\documentclass[preprint,12pt]{elsarticle}

\usepackage{amsfonts,amssymb,amsbsy,textcomp,marvosym,amsmath,caption,threeparttable,amsthm,subfigure,float,lastpage,lscape}
\usepackage{eurosym,mathrsfs,fancyhdr,CJK,multicol,graphics,indentfirst,color,bm,upgreek,booktabs,graphicx,multirow,warpcol}
\usepackage{url}
\usepackage{tikz}
\usepackage{adjustbox}
\usepackage{booktabs}

\usepackage{fixme}
\usepackage{xspace}
\usepackage{xcolor}

\newtheorem{example}{Example}

\def\graph{\ensuremath{G}}

\def\vertices{\ensuremath{V}}
\def\edges{\ensuremath{E}}
\def\features{\ensuremath{X}}
\def\neighbors{\ensuremath{N}}

\def\gtrain{\ensuremath{\graph_{\mathit{train}}}}
\def\vtrain{\ensuremath{\vertices_{\mathit{train}}}}
\def\etrain{\ensuremath{\edges_{\mathit{train}}}}
\def\vtest{\ensuremath{\vertices_{\mathit{test}}}}

\def\gtrainneg{\ensuremath{\gtrain^-}}

\newcommand\posneighbors[1]{\ensuremath{\neighbors^+_{#1}}}
\newcommand\negneighbors[1]{\ensuremath{\neighbors^-_{#1}}}

\newcommand\myexample[2]{\ensuremath{\langle #1,#2\rangle}}

\def\minPure{\ensuremath{\mathit{minPure}}}
\def\maxSpurious{\ensuremath{\mathit{maxSpurious}}}
\newcommand\purity[2]{\ensuremath{[#1,#2]}}

\def\ogblc{\texttt{ogbl-citation2}\xspace}
\def\sample{\texttt{sample}\xspace}
\def\sampleu{\texttt{sample\_u}\xspace}
\def\ba{\texttt{ba}\xspace}
\def\ego{\texttt{ego}\xspace}
\def\er{\texttt{er}\xspace}
\def\sampleba{\texttt{sample+ba}\xspace}

\newcommand\nemb[1]{\ensuremath{\mathbf{h}_{#1}}}
\newcommand\eemb[1]{\ensuremath{\mathbf{x}_{#1}}}

\newcommand{\orcid}[1]{\textsuperscript{\href{https://orcid.org/#1}{\includegraphics[scale=0.04]{00_orcid.pdf}}}}

\journal{Information Sciences}

\begin{document}

\begin{frontmatter}



\title{Introducing New Node Prediction in Graph Mining: Predicting All Links from Isolated Nodes with Graph Neural Networks}

\cortext[mycorrespondingauthor]{Corresponding author}
\author[dia]{Damiano Zanardini}\ead{damiano.zanardini@upm.es}
\author[dia]{Emilio Serrano\corref{mycorrespondingauthor}}\ead{emilio.serrano@upm.es}

\address[dia]{Departamento de Inteligencia Artificial, ETSI Informáticos, Universidad Politécnica de Madrid, 28660 Boadilla del Monte, Madrid, Spain}


\begin{abstract}
This paper introduces a new problem in the field of graph mining and social network analysis  called \emph{new node prediction}. More technically, the task can be categorized as \emph{zero-shot out-of-graph all-links prediction}.  This challenging problem aims to predict all links from a new, isolated, and unobserved node that was previously disconnected from the graph.  Unlike classic approaches to link prediction (including \emph{few-shot} out-of-graph link prediction), this problem presents two key differences: (1) the new node has no existing links from which to extract patterns for new predictions; and (2) the goal is to predict not just one, but all the links of this new node, or at least a significant part of them.  Experiments demonstrate that an architecture based on Deep Graph Neural Networks can learn to solve this challenging problem in a bibliographic citation network.
\end{abstract}



\begin{keyword}
node prediction \sep link prediction \sep graph Neural Networks \sep graph mining  \sep deep learning
\end{keyword}

\end{frontmatter}



\section{Introduction}

Graphs and network data are essential components in modeling many complex systems in a wide variety of fields such as social network analysis, biology, transportation, and the World Wide Web \cite{ZhouEtAl2020}. The study of graph and networks can reveal crucial yet hidden information such as communities or clusters, influence spread, and central entities (e.g., hubs or bridges in a network). The use of graphs is also at the heart of some of the most promising proposals for the Semantic Web: \textit{Knowledge Graphs} \cite{DBLP:journals/corr/NickelMTG15}.

\textit{Link prediction} is one of the main problems in graph mining. This task aims to predict new interactions or relationships between entities, for example potential friendships in social networks or interactions between proteins in biological networks. Link prediction models can also help to complete knowledge graphs, which are often incomplete due to their large-scale and the expense of data acquisition.

This paper introduces a novel learning task in graph mining, the problem of ``new node prediction''. This task is a variant of the link prediction problem, but focuses on predicting links for new nodes that do not possess existing patterns. This problem can be more technically defined as \emph{zero-shot out-of-graph all links prediction}. The two crucial distinctions to classical link prediction approaches are: (1) the node is brand new with no existing links to extract patterns for new predictions, and (2) the objective is to predict not just a single, but all the links of this new node, or a substantial portion thereof. 

Similarly to other zero-shot learning scenarios, auxiliary information may be employed to encode observable distinguishing properties of the new nodes. In this paper,  experiments are conducted using a bibliographic citation network, where the title and abstract of the articles serve as auxiliary information for the new node prediction problem. Experiments demonstrate that Deep Learning architectures are capable of predicting a large percentage of links (citations) for new nodes (articles) based on the observed citation graph and the auxiliary information provided. 

More specifically, this paper proposes variants of \textit{Deep Graph Neural Networks} (GNNs) to solve this new and challenging problem. 
In recent years, GNNs have emerged as a key technology for predicting network data. GNNs provide an effective way to filter relevant structural information in a network through their non-linear operations. They have been successful in predicting links in various contexts, from social networks to protein-protein interactions and citation graphs. 

This paper contributes with: (1)  the introduction and formalization of the ``new node prediction'' task; (2) the proposal of a GNN-based architecture to learn it; (3) the experimental results that demonstrate the feasibility of solving this task in the context of a bibliographic citation graph; and finally, (4) the trained GNN models (for this task and this network), which can be used to recommend cites in unpublished papers.

The remainder of this paper is structured as follows. After discussing the related works and some preliminaries in sections \ref{rw} and \ref{preliminaries}, section \ref{sec:theProblemOfNewNodePrediction} introduces the novel problem of ``new node prediction''.  Section \ref{arc} describes the neural network architecture proposed to learn this task.  Section \ref{exp} addresses the experimental results to study the feasibility of the problem at hand. Section \ref{con} concludes and gives the future work.




\subsection{Related work} \label{rw}
 
\emph{Link prediction} \cite{LibenNowellKleinberg2007} is a task in graph and network analysis that aims to predict future or missing connections between nodes in a network. Given a partially observed network, link prediction involves inferring which links are most likely to be added based on the observed connections and the network structure. The two basic approaches for link prediction are: (1) \emph{topology-based methods}, and (2) \emph{node attribute-based methods}. Topology-based methods, such as Common Neighbors and Jaccard Measure, assume that nodes with similar network structures are more likely to form a link. Node attribute-based methods, such as Euclidean Distance and Cosine Similarity, predict the existence of a link based on the similarity of node attributes.

Many current approaches for link prediction combine attribute and topology-based methods. Some of these mixed methods include: (1) \emph{ node representation / graph feature-based methods} (such as DeepWalk \cite{deepwalk} and node2vec \cite{node2vec}); (2) \emph{Knowledge Graph Embeddings} (KGEs) (such as TransE, DistMult, ComplEx, and ConvE) \cite{KGEREVIEW}; and (3) \emph{Graph Neural Networks} (GNNs) (such as Graph Convolutional Networks and Graph Attention Networks) \cite{DBLP:conf/nips/ZhangC18}.

GNNs achieve high accuracy in link prediction by jointly leveraging graph topology and node attributes. When considering the popular Citeseer dataset\footnote{\url{https://relational.fit.cvut.cz/dataset/CiteSeer}} for testing link prediction, some GNNs-based methods report an AUC (Area under the ROC Curve) over 96\%, including the use of \emph{graph autoencoders} (GAEs) \cite{NESS} and \emph{variational graph autoencoders} (VGAEs) \cite{Seong2021}. Different methods to improve GNNs learning, such as incorporating clustering information \cite{infoClust} or utilizing new pooling schemes \cite{walkPooling}, also achieve outstanding results. Although we experiment with GNNs, this paper does not focus on a specific architecture or method for predictions: instead, it defines a new related problem and investigates its feasibility.

There are many domains and types of graphs in which including all entities during the training process is impractical or unfeasible. For example, real-world \emph{knowledge graphs} (KG) involve the constant occurrence of new entities. For this reason, there are several works and research lines in the literature that address link prediction in scenarios where little information is known about the connectivity of new entities. Wang et al.~\cite{wang2020logic} consider the problem of prediction of links in a new emerging node, which also brings new links to the existing (and observed) nodes in a KG.  Baek et al.~\cite{baek2020learning} also study essentially the same problem under the very descriptive term of \emph{transductive few-shot out-of-graph link prediction}, obtaining significant results on benchmark datasets for knowledge graph completion and prediction of drug-drug interaction. Before these works, Hamaguchi  et al.~\cite{Hamaguchi2017} studied the problem of \emph{out-of-knowledge-base (OOKB) entity in knowledge-base completion (KBC)}. This task requires answering queries regarding test nodes that were not observed at training time. This challenging problem can be seen as a \emph{one-shot out-of-Graph link prediction}, i.e. predicting new links to a new node that emerges with just one link to observed nodes. Bose et al.~\cite{Bose2019} also address the few-shot link prediction pointing out that most link prediction techniques perform poorly on this task.  Jambor et al.~\cite{Jambor2021} empirically explore the limits of existing few-shot Link Prediction methods, concluding that having only a few examples of a relation fundamentally limits models from using fine-grained structural information.

Recently, Seong and Kim \cite{Seong2021} took the link prediction problem to another level. They found that GAEs and VGAEs performed poorly when dealing with nodes whose degree is zero or isolated nodes. This scenario is more realistic than it seems. The authors gave a very illustrative example of a social network of students where students are the nodes and the links indicate a ``friendship'' relationship. In this scenario, the task of ``finding friends of students'' is a link prediction problem.  What they found was that most link prediction methods could not be used for ``freshmen'', i.e., new nodes that do not have any connections. In these isolated nodes, the content of the features (e.g., the circles or the hobby of the students) plays an important role in this link prediction problem because there is no connectivity information. Their experimental results showed that the content features of isolated nodes tend to go to zero in various graph-structured networks, unlike their proposed \emph{Graph Normalized AutoEncoder} (VGNAE). 

Similarly to Seong and Kim \cite{Seong2021}, we deal with link prediction on isolated nodes. However, we want to explore the possibility of predicting all (or a significant amount of) links of a previously isolated  (non-existing) node. To the best of our knowledge, this is a problem that has not been explored previously in the specialized literature. Knowing all links of a node means, in a number of graphs, knowing all the relevant information of this node. Therefore, we believe that the term \emph{new node prediction} is more appropriate than \emph{link prediction}. However, this can also be seen as \emph{zero-shot out-of-graph all-links prediction}. This new problem is not just an aggregation of link prediction with isolated nodes. Even if a model can predict a large number of new links of isolated nodes, this same model can be incapable of predicting most of the links of a specific isolated node. Another difference with respect to related work \cite{Seong2021} is that, admitting the great importance of feature content in the prediction of isolated node links, their article does not specify what these are, nor experiments with different initial features. 

Wang et al.~\cite{KGEREVIEW} detail research works which incorporate additional information for link prediction; among others: node types, link types, relation paths, logic rules, and textual descriptions. As in any zero-shot learning setting, the use of some form of auxiliary information is essential. We experiment with feature content based on textual description embeddings among other alternatives, evaluating their impact on the novel problem of \emph{new node prediction}. 

 
There are other problems related with the task of \emph{new node prediction} introduced here.  Galkin et al.~\cite{galkin2022open} provides benchmarks for inductive link prediction on KGs. In inductive settings, the training graph and the network where links must be predicted are different. The \emph{inductive inference graph} can contain a combination of seen and unseen nodes (i.e., the setting is semi-inductive) or only unseen nodes (i.e., fully inductive), and either the same or a subset of links from the training graph. Although this setup also makes the link prediction problem more challenging, the ability to predict all links for an emerging and isolated node is not evaluated. Song et al.~\cite{Song2022} define the problem of \emph{graph learning with out-of-distribution nodes}, which means detecting nodes that do not belong to the known distribution. Hence, this task can be seen as a type of node classification instead of a case of link prediction. Zhang  et al.~\cite{ZhangDL0YC022} comprehensively surveys work on few-shot learning on graphs.  In the mentioned survey no zero-shot task has been explored except zero-shot relation prediction \cite{Qin2020}. This supports the novelty of the presented \emph{ new node prediction} task, which is a case of zero-shot learning.

\subsection{Preliminaries} \label{preliminaries}

A \emph{graph} $\graph = (\vertices,\edges,\features)$ is made of a set \vertices of \emph{nodes}/\emph{vertices}, and a set $\edges \subseteq \vertices \times \vertices$ of \emph{edges} between them.  Graphs can be either \emph{directed} or \emph{undirected}; an undirected graph is such that connections are symmetric, that is, for every $(n_1,n_2) \in \edges$, the symmetric pair $(n_1,n_2)$ also belongs to $\edges$; on the other hand, in a directed graph, there can be an edge from $n_1$ to $n_2$ even though there is no edge from $n_2$ to $n_1$. A directed edge is often called a \emph{link}.
Optionally, edges can have \emph{weights} indicating the strength of a connection, and a vertex can have characteristics (\emph{features}) representing additional information: $\features$ is a matrix of features on $\vertices$.
Given a node $n$, the set $\neighbors(n)$ of its \emph{neighbors} is the set of vertices directly connected to $n$; if the graph is undirected, then $\neighbors(n) = \{m~|~(m,n) \in \edges\}$; on the other hand, in a directed graph, we can distinguish between \emph{incoming} neighbors (vertices from which there is a direct link to $n$), \emph{outgoing} neighbors, or the union of both sets.
Given a graph $\graph = (\vertices,\edges)$, the \emph{sub-graph} $\graph'$ of $\graph$ \emph{induced} by $\vertices' \subseteq \vertices$ is such that $\graph' = (\vertices',\edges')$ where $\edges' = \{(n_1,n_2)~|~n_1,n_2 \in \vertices'\}$.

\section{The problem of new node prediction}
\label{sec:theProblemOfNewNodePrediction}


The task of \emph{new node prediction} consists of learning new nodes from an existing graph.
Unlike link prediction, where the set of nodes is fixed and a new link can be easily characterized by the existing nodes it links, new nodes in a graph have to be identified by their connections (new edges/links) with existing nodes.  The present paper does not investigate the problem of learning features for new vertices.

At training time, the model is supposed to learn \emph{examples} (nodes with their neighbors); i.e., to tell positive (actually existing in the graph) from negative examples.  At test time, new (previously unseen) examples are provided, and binary classification is run on them.
A test example is previously unseen because it represents the edges (i.e., the neighbors) of a node that was not available at training time; however, its neighbors may belong to the training information, as usual in graph analysis, where training and test data are not completely independent.

More concretely, the graph nodes $\vertices$ of a graph $\graph = (\vertices,\edges)$ are divided into \emph{training} nodes ($\vtrain$) and \emph{test} nodes ($\vtest$).
The \emph{training graph} $\gtrain = (\vtrain,\etrain)$ induced by $\vtrain$ is made of the training nodes and all edges between them.  At training time, \emph{positive and negative examples} are created for binary classification, as follows.
Let $\etrain^-$ be the set of \emph{negative edges}; such a set is generated once and for all before training: $\etrain^-$ are the edges of the graph $\gtrainneg$ that has the same nodes as $\gtrain$, but only edges that do not exist in $\gtrain$. More formally: $\gtrainneg$ has nodes $\vtrain^- = \vtrain$ and edges $\etrain^- \subseteq (\vtrain \times \vtrain) \setminus \etrain$.
For each target node $t$, let $\posneighbors{t}$ (resp., $\negneighbors{t}$) be the \emph{positive} (resp., \emph{negative}) \emph{neighbors}, i.e., the set of nodes $s$ that are connected to $t$ by positive (resp., negative) edges $(s,t)$.
Negative edges are produced randomly, and, if possible, their number is comparable to the number of positive edges; however, this may depend on each specific graph. 

The first definition that comes to mind for examples is that a positive example should only allow positive edges, and include all of them, and a negative example should only allow negative edges, and include all of them. However, a certain degree of \emph{spuriousness} or \emph{noise} can be allowed in order to make the definition of the problem more flexible: given an example, let \emph{pure} edges be those edges with the same \emph{polarity} as the example itself (i.e., positive edges if the example is positive and negative edges if the example is negative); on the other hand, let \emph{spurious} edges be those edges with different polarity w.r.t.~the example.
The parameters $\minPure$ and $\maxSpurious$ determine how examples are generated:
\begin{itemize}
\item $\minPure \in [0,100]$ determines the \emph{minimum} percentage of pure edges that have to be included in an example;
\item $\maxSpurious \in [0,100]$ determines the \emph{maximum} percentage of spurious edges that have to be included in an example.
\end{itemize}
In the following, \emph{purity} will correspond to the pair $\purity{\minPure}{\maxSpurious}$.


Therefore, a \emph{positive train example} comes to be a pair $\myexample{S^+\cup S^-}{t}$ where
    \begin{itemize}
        \item $S^+$ contains \emph{at least} $(\minPure)\%$ of the nodes in $\posneighbors{t}$;
        \item $S^-$ contains \emph{at most} $(\maxSpurious)\%$ of the nodes in $\negneighbors{t}$; and
        \item $t$ is a node in $\gtrain$.
    \end{itemize}
A \emph{negative train example} is defined accordingly: it is a pair $\myexample{S^-\cup S^+}{t}$ where
    \begin{itemize}
        \item $S^-$ contains \emph{at least} $(\minPure)\%$ of the nodes in $\negneighbors{t}$;
        \item $S^+$ contains \emph{at most} $(\maxSpurious)\%$ of the nodes in $\posneighbors{t}$; and
        \item $t$ is a node in $\gtrain$.
    \end{itemize}
In other words, an example is generally made with \emph{pure} edges (i.e., positive edges if the example is positive, or negative edges if the example is negative), but also a usually smaller portion of \emph{spurious} edges (i.e., negative edges if the example is positive or positive edges when the example is negative).
To include spurious edges in examples is a way to make the problem more general, because the model can learn sets of edges even in the presence of some amount of noise (as shown in Section \ref{sec:ComparisonOnDifferentValuesForTrainAndTestPurity}).

Positive and negative training examples are randomly generated, provided that purity conditions are met; afterwards, they are labeled with $1$ and $0$, respectively, and a binary classifier is run on them.  In the experiments, exactly one positive and one negative example is generated for each node; however, this could be easily modified.

Therefore, the training phase is meant to teach the model how to tell positive from negative examples that are included in the training graph $\gtrain$ or its negative counterpart $\gtrainneg$.

For the \emph{testing} of the model, the test examples are created similarly, the only difference being that the target node $t$ is a test node instead of a training node. Positive and negative edges are referred, in this case, to the whole $\graph$ and $G^-$ instead of just $\gtrain$ and $\gtrainneg$.
In other words, the trained model is supposed to be able to predict whether or not a (previously unseen) test node is likely to have some specific set of links to other nodes.
To this end, allowing spurious edges in test examples makes the prediction problem more difficult due to the noise in test data. Note that the parameters $\minPure$ and $\maxSpurious$ do not need to be the same at training and test time.
Section \ref{sec:ComparisonOnDifferentValuesForTrainAndTestPurity} gives some insight into how purity affects experimental results.
 
\begin{figure}[h]
  \centering
  \[
  \begin{array}{ccc}
    \begin{tikzpicture}
      \node (n1) at (0,0) {$n_1$};
      \node (n2) at (1.2,.5) {$n_2$};
      \node (n3) at (1.2,-0.7) {$n_3$};
      \node (n4) at (2.4,0) {$n_4$};
      \node (n5) at (-.6,-1) {$n_5$};
      \node (n6) at (0.5,-2) {$n_6$};
      \node (n7) at (2.4,-2) {$n_7$};
      \node (n8) at (3.2,-1) {$n_8$};
      
      \draw (n1) -- (n2);
      \draw (n1) -- (n6);
      \draw (n2) -- (n4);
      \draw (n3) -- (n4);
      \draw (n3) -- (n5);
      \draw (n3) -- (n6);
      \draw (n3) -- (n7);
      \draw (n4) -- (n7);
      \draw (n6) -- (n8);
      \draw (n7) -- (n8);
    \end{tikzpicture}
    &
    \begin{tikzpicture}
      \node (n1) at (0,0) {$n_1$};
      \node (n2) at (1.2,.5) {$n_2$};
      \node (n3) at (1.2,-0.7) {$n_3$};
      \node (n4) at (2.4,0) {$n_4$};
      \node (n5) at (-.6,-1) {$n_5$};
      \node (n6) at (0.5,-2) {$n_6$};
      
      \draw (n1) -- (n2);
      \draw (n1) -- (n6);
      \draw (n2) -- (n4);
      \draw (n3) -- (n4);
      \draw (n3) -- (n5);
      \draw (n3) -- (n6);
    \end{tikzpicture}
    &
    \begin{tikzpicture}
      \node (n1) at (0,0) {$n_1$};
      \node (n2) at (1.2,.5) {$n_2$};
      \node (n3) at (1.2,-0.7) {$n_3$};
      \node (n4) at (2.4,0) {$n_4$};
      \node (n5) at (-.6,-1) {$n_5$};
      \node (n6) at (0.5,-2) {$n_6$};
      
      \draw[dashed] (n1) -- (n3);
      \draw[dashed] (n1) -- (n4);
      \draw[dashed] (n1) -- (n5);
      \draw[dashed] (n2) -- (n3);
      \draw[dashed] (n2) -- (n5);
      \draw[dashed] (n2) -- (n6);
      \draw[dashed] (n4) -- (n5);
      \draw[dashed] (n4) -- (n6);
      \draw[dashed] (n5) -- (n6);
    \end{tikzpicture}\\[3mm]
    \mbox{(a)} & \mbox{(b)} & \mbox{(c)}\\[3mm]
    \begin{tikzpicture}
      \node[draw] (n1) at (0,0) {$n_1$};
      \node (n2) at (1.2,.5) {$n_2$};
      \node (n3) at (1.2,-0.7) {$n_3$};
      \node (n4) at (2.4,0) {$n_4$};
      \node (n5) at (-.6,-1) {$n_5$};
      \node (n6) at (0.5,-2) {$n_6$};
      
      \draw (n1) -- (n2);
      \draw (n1) -- (n6);        
    \end{tikzpicture}
    &
    \begin{tikzpicture}
      \node (n1) at (0,0) {$n_1$};
      \node (n2) at (1.2,.5) {$n_2$};
      \node[draw] (n3) at (1.2,-0.7) {$n_3$};
      \node (n4) at (2.4,0) {$n_4$};
      \node (n5) at (-.6,-1) {$n_5$};
      \node (n6) at (0.5,-2) {$n_6$};
      
      \draw (n1)[dotted] -- (n3);
      \draw (n3) -- (n4);
      \draw (n3) -- (n5);
      \draw (n3) -- (n6);
    \end{tikzpicture}
    &
    \begin{tikzpicture}
      \node (n1) at (0,0) {$n_1$};
      \node (n2) at (1.2,.5) {$n_2$};
      \node (n3) at (1.2,-0.7) {$n_3$};
      \node[draw] (n4) at (2.4,0) {$n_4$};
      \node (n5) at (-.6,-1) {$n_5$};
      \node (n6) at (0.5,-2) {$n_6$};
      
      \draw (n2)[dotted] -- (n4);
      \draw (n4) -- (n5);
      \draw (n4) -- (n6);        
    \end{tikzpicture}\\
    \mbox{pos.:}~(\{n_2,n_6\},\mathbf{n_1}) & \mbox{pos.:}~(\{\underline{n_1},n_4,n_5,n_6\},\mathbf{n_3}) & \mbox{neg.:}~(\{\underline{n_2},n_5,n_6\},\mathbf{n_4}) \\[3mm]
    & \mbox{(d)}
    
  \end{array}
  \]
  
  \caption{(a) A tiny graph $G$ with 8 nodes; (b) the training portion $\gtrain$ of it: $\{n_1,n_2,n_3,n_4,n_5,n_6\}$ are the training nodes, whereas $\{n_7,n_8\}$ are the test nodes; (c) the graph $\gtrainneg$ of negative edges; (d) some (positive or negative) training examples with $\minPure = 0.7$ and $\maxSpurious = 0.5$ (target nodes are boxed; spurious edges are dotted).}
  \label{fig:posNegExamples}
\end{figure}
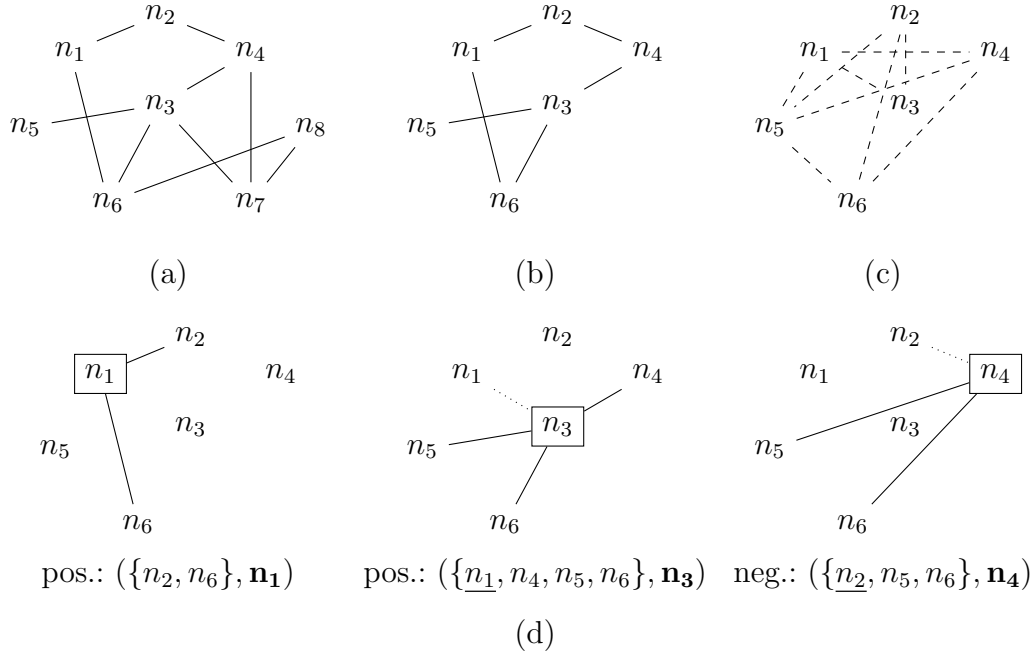

\begin{example}
    Consider the graph in Figure \ref{fig:posNegExamples} (a), and the node partition $\vtrain = \{n_1,n_2,n_3,n_4,n_5,n_6\}$ and $\vtest = \{n_7,n_8\}$.
    Node $n_3$ has four neighbors: $\posneighbors{n_3} = \{n_4,n_5,n_6,n_7\}$; however, only $\{n_4,n_5,n_6\}$ is the set of positive neighbors after computing the subgraph induced by $\vtrain$ (b).
    Moreover, once the negative edges have been generated randomly (c), $\negneighbors{n_3}$ becomes $\{n_1,n_2\}$.
    The most direct choice for a positive training example on $n_3$ is to include all training positive edges, and no negative edges: the example would be $\myexample{\{n_4,n_5,n_6\}}{n_3}$, and would correspond to $\minPure = 100\%$ and $\maxSpurious = 0\%$, i.e., a purity of \purity{100}{0}.

    However, different choices for these two parameters allow the generation of more flexible examples. Let $\minPure = 70\%$ and $\maxSpurious = 50\%$, corresponding to purity \purity{70}{50}: in this case, for any positive example $\myexample{S}{n_3}$, $S$ will have to contain at least $3 = \lceil 3 \cdot 70\%\rceil$ nodes from $\posneighbors{n_3}$ (there are three such nodes), and at most $1 = \lfloor 2 \cdot 50\%\rfloor$ nodes from $\negneighbors{n_3}$.  $S$ will be chosen randomly according to these conditions: for example, $\{n_1,n_4,n_5,n_6\}$ (c) is a legal set, where three pure edges $n_4$, $n_5$ and $n_6$ have been selected, and only one spurious edge $n_1$.
\end{example}

\section{The Deep Learning architecture for Node Prediction}
\label{arc}

This section describes the neural network architecture and explains how the information contained in the graph is processed.


The main components of the neural architecture are
\begin{itemize}
    \item A \emph{Graph Neural Network} (GNN) \cite{KipfWelling2017,ZhouEtAl2020} that generates a \emph{node embedding} $\nemb{n} \in \mathbb{R}^k$ for every node $n$ and dimensionality $k$. Experiments report on the use of different GNN architectures.
    \item An \emph{example generator} that produces positive and negative examples, both in training and test time.
    \item A \emph{neural predictor} that computes an embedding for each example and applies binary classification on it.
\end{itemize}

From node embeddings, embeddings of examples are computed.  The embedding $\eemb{e}$ of an example $e = \myexample{S}{t}$ is the vector resulting from the concatenation of the average vector $\nemb{S}$ of all the embeddings $\nemb{n}$ for $n\in S$, together with the embedding $\nemb{t}$:
\[ \eemb{\myexample{S}{t}} = (\mathit{average}(\{ \nemb{n}~|~n \in S \})) \circ \nemb{t} \]
The predictor is a Multilayer perceptron (MLP)  with a configurable number of dense layers and artificial neurons using the rectified linear unit (ReLU) activation function. The dimensionality of the input to the first layer
is the number of dimensions of the example embedding $\eemb{e}$, which is $2k$, that is, twice the number of dimensions of the embeddings of the nodes $\nemb{n}$.  The \emph{sigmoid} function is applied to the result of the last layer for binary classification.
The GNN and predictor parameters are updated jointly at training time, consistently with \emph{supervised learning} tasks with Graph Neural Networks.

\begin{figure}
    \centering
    \resizebox{8cm}{5.5cm}{
    \begin{tikzpicture}
        \node[draw,minimum height=5cm,minimum width=3.0cm] (gnn) at (1.5,-3.4) {};
        \node (graph) at (2.5,-7) {graph};
        \node (nf) at (0.5,-7) {node features};
        \draw[->] (graph) -- (gnn);
        \draw[->] (nf) -- (gnn);

        \node[draw,minimum height=0.6cm,minimum width=2.6cm] (relu2) at (1.5,-1.4) {ReLU, norm.};
        \node[draw,minimum height=0.6cm,minimum width=2.6cm] (gnn2) at (1.5,-2.4) {GNN layer};
        \node[draw,minimum height=0.6cm,minimum width=2.6cm] (relu1) at (1.5,-4.4) {ReLU, norm.};
        \node[draw,minimum height=0.6cm,minimum width=2.6cm] (gnn1) at (1.5,-5.4) {GNN layer};

        \draw[->] (gnn1) -- (relu1);
        \draw[dotted,->] (relu1) -- (gnn2);
        \draw[->] (gnn2) -- (relu2);

        \node[draw] (generator) at (5,-4) {$\begin{array}{c}
        \mbox{generator} \\ \mbox{of labeled} \\
        \mbox{examples} \\ (S,t) \end{array}$};

        \draw[->] (graph) -- (generator);

        \node[draw,minimum height=0.6cm] (avg) at (4.5,-1.8) {AVG};

        \node[draw,minimum height=0.6cm] (concat) at (5,-0.25) {concat};
        \draw[->] (avg) -- (concat);

        \draw[->] (generator) -- node[left]{$S$} (avg);
        \draw[->] (generator) edge[bend right=20] node [right] {$t$} (concat);

        \draw[->] (gnn) .. controls (2,0) and (3,0) .. node[above]{node embs. $\nemb{} \in\mathbb{R}^k$} (concat);
        \draw[->] (gnn) .. controls (2,0) and (3,0) .. (avg);

        \node[draw,minimum height=1.6cm,minimum width=2.0cm] (predictor) at (8,-2) {predictor};
        \draw[->] (concat) edge[bend left=20] node[auto]{example embs. $\eemb{(S,t)} \in\mathbb{R}^{2k}$} (predictor);

        \node[draw,minimum height=0.6cm,minimum width=1.2cm] (sigma) at (8,-4) {$\sigma$};
        \draw[->] (predictor) -- (sigma);

        \node[draw,minimum height=0.6cm] (correct) at (8,-5.5) {correct?};
        \draw[->] (sigma) -- (correct);
        \draw[->] (generator) edge[bend left=30] node[left]{label} (correct);

        \node (y) at (7,-7) {yes};
        \node (n) at (9,-7) {no};

        \draw[->] (correct) -- (y); 
        \draw[->] (correct) -- (n); 
    \end{tikzpicture}}
    
    \caption{A Deep Learning architecture for the new node prediction problem.}
    \label{fig:architecture}
\end{figure}
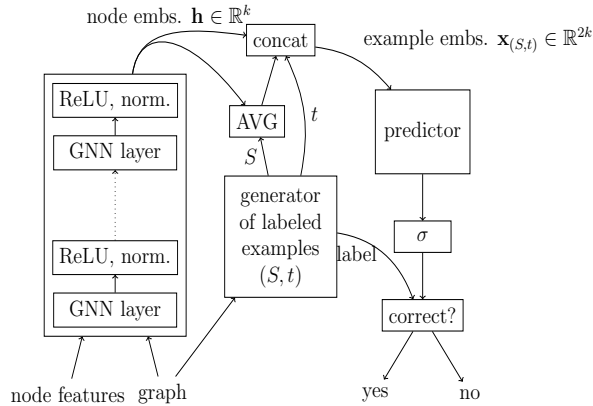

Figure \ref{fig:architecture} shows the neural architecture in detail. Note that the task of ``new node prediction'' can be learned using different neural architectures; this aspect is left for future work.

\section{Experiments}
\label{exp}

This section details the experiments to demonstrate the feasibility of the ``new node prediction'' task introduced in this paper. The datasets used, the experimental setup, and the results obtained are described.

\subsection{Datasets}


The architecture described previously is used with several synthetic graphs and with a subset of the \ogblc graph\footnote{\url{https://ogb.stanford.edu/docs/linkprop/\#ogbl-citation2}}.

The \ogblc dataset is a directed, homogeneous graph representing the citation network between a subset of papers extracted from the Microsoft Academic Graph \cite{WangEtAl2020}. Each node is a paper with 128-dimensional features that summarizes its title and abstract, and each directed edge indicates that one paper cites another.
Due to the size of the original graph (almost three million nodes and several million edges), experiments are carried out on subgraphs, called \sample in the following, whose node number is in the thousands.
To avoid excessive sparseness due to a random choice of nodes, the highest-degree nodes are chosen.  For example, the average degree of the sample graph with ten thousand nodes is $7.99$, compared to $10.38$ in the original graph.

The task that is usually performed on \texttt{ogbl-citation2} is \emph{link prediction}, which in this context means learning missing citations. In contrast, the contribution presented in this paper attempts to  \emph{learn new papers}, where a new paper is characterized by the papers it cites. To this end, edges have been reversed, ie. the relation they represent becomes \emph{is-cited-by} instead of \emph{cites}. Therefore,  the connectivity of a new node/paper corresponds to the existing papers to which it cites/is-connected.  This is meaningful in the context of node prediction, since \ogblc is acyclic, so there is a clear order between existing and new nodes (new articles by which existing papers are cited).
In the following, $[\cdot]_n$ will denote a graph with $|\vertices| = n$ nodes (for example, $\sample_{10K}$ indicates the sample of \ogblc with ten thousand nodes, although the subscript will often be dropped for clarity).

In general, our definition of the learning task works for both directed and undirected graphs.  In fact, we also run experiments on the undirected version \sampleu of \sample, together with several randomly generated graphs:
\begin{itemize}
    \item An undirected \emph{egocentric} network $\ego$ that is the $k$-hop sub-graph starting from a randomly-chosen node in the pre-computed undirected version of \ogblc; $k$ is the number of hops that are needed to reach the intended number $n$ of nodes: for the sake of experiments, the graph is randomly generated many times until the number of nodes for a given $k$ is close enough to $n$ (between $2n/3$ and $3n/2$).
    \item A Barab\'asi-Albert \cite{Barabasi99emergenceScaling} graph \ba with the specified number of nodes.
    \item A Erd\"os-Renyi \cite{erdos59a} graph \er with the specified number of nodes.
    \item A \sample graph whose number of nodes is equal to the intended number $|\vtrain|$ of train nodes, to which the Barab\'asi-Albert algorithm has been applied, giving the \sampleba graph with $|\vtrain|+|\vtest|$ nodes.
\end{itemize}

Regarding the partition into training and test data, the graph nodes are randomly divided into training ($80\%$) and test nodes ($20\%$), and examples are generated as described in Section \ref{sec:theProblemOfNewNodePrediction}.  In the case of \sampleba, the partition is clearly not random; instead, the set of test nodes is exactly the set of nodes generated by the Barab\'asi-Albert model, which become the $20\%$ of the total nodes.

\subsection{Experimental setup}

The model\footnote{Code is available at \url{https://github.com/damianozanardini/nodePrediction/} to allow the interested researcher to use and extend our work.} has been implemented by using PyTorch 2.0.1 and PyTorch Geometric 2.4.0.
Experiments have been carried out on a ``Pay As You Go'' subscription to Google Colab \cite{Bisong2019}, using the A100 GPU. A variety of standard GNN architectures have been used, namely GCN \cite{KipfWelling2017}, ClusterGCN \cite{ChiangEtAl2019}, GAT \cite{VelickovicEtAl2018}, and SAGE \cite{HamiltonEtAl2017}.

It is not possible to report on all the choices for the hyper-parameters.  We only show the results of some significant experiments that can provide some insight into how the algorithm performs under different conditions.
GNN architectures have been defined with five layers, and they compute node embeddings with dimensionality 128, which seems to be enough for the size of the graphs that we take into account (experiments have been carried out with different embedding sizes, with no significant improvements).
Data are fed to the Graph Neural Networks in batches of predefined size, proportional to the size of the graph; in turn, the number of nodes in the graph determines the number of positive and negative examples in the dataset.

Unless otherwise specified, the maximum number of epochs is 50; \emph{early stopping} is performed (with a default value for \emph{patience} of 8 epochs) in a validation set that is 10\% of the train set.
That is, if the loss in the validation set at epoch $i$ is better than the loss at epochs $i{+}1,...,i{+}8$, then training is aborted, and the parameter values at $i$ are kept instead of the newer ones.

The following tables in Section \ref{sec:results} report on the experimental results. The meaning of each column is as follows:
\begin{itemize}
    \item \texttt{\#nodes} is $|\vertices|$, i.e., the number of nodes in the graph;
    \item \texttt{GNN} is the GNN architecture (GCN, ClusterGCN, GAT, SAGE);
    \item \texttt{train purity} and \texttt{test purity} is the purity of train and test examples, respectively; in most examples, intermediate values for purity are used, namely, \purity{80}{10} for both train and test examples; however, see Section \ref{sec:ComparisonOnDifferentValuesForTrainAndTestPurity} for a discussion on how these values affect the precision of the model;
    \item \texttt{lr} is the \emph{learning rate} (unless differently specified, $\texttt{lr}=0.001$) of the \texttt{torch.optim.Adam} optimizer \cite{KingmaB14};
    \item \texttt{train time} is the training time in seconds; this does not include the time for generating positive and negative examples, that is, however, much smaller than the proper training time;
    \item \texttt{accuracy}, or \texttt{acc.}, is the percentage of examples that are classified correctly at test time (mean of five executions of the test phase on different, randomly-generated sets of test examples);
    \item \texttt{hits@k}: the \emph{Hits@k} metric for different values of $k$ (mean of five executions of the test phase); the number of negative examples among which each positive example is ranked is given in each table;
    \item \texttt{MRR} is the \emph{mean reciprocal rank} (mean of five executions of the test phase).
\end{itemize}

\subsection{Results}
\label{sec:results}

The following sections report experimental results on different learning configurations.
Although not all the possible configurations have been taken into account, experiments have been designed in order to provide insights on the nature of the problem and the best choices for options and hyper-parameters.

In general, the figures seem to heavily depend on how train and test examples are generated. Since example generation has a randomized component, most results are computed as the average of a few runs; however, more trustworthy results would be obtained by increasing the number of runs.

\subsubsection{Comparison on different GNN architectures}

We first present results on \sampleu by using different GNN architectures (Table \ref{tab:gnntype}).
In each line, the figures are computed as the average of three runs.
In some runs, the combination of the graph size, the GNN architecture, and the learning rate made learning infeasible: the model parameters did not stabilize, and all test examples were classified as the same class.  In these cases, the table specifies the number of meaningful runs on which averages are computed: for example, with ClusterGCN with learning rate 0.01, only two runs were meaningful; moreover, no runs were meaningful with learning rate 0.05, so that the corresponding table row is left blank.
The reason for not considering this kind of meaningless runs is that figures usually do not make any sense: for example, it is often the case that the accuracy gets stuck around 50\%, which is understandable given the situation, but the values for MRR are close to 1, which may be wrongly considered as a very good result; therefore, we choose not to consider these numbers when computing averages.

\noindent
\begin{table}[h!]
\begin{adjustbox}{max width=\textwidth}
\begin{tabular}{cccccccccccccc}
    \toprule
    & & & \textbf{train} & \% & \multicolumn{8}{c}{\texttt{hits@k} (out of 20\% neg. examples)} & \\
    \texttt{\#nodes} & \textbf{GNN} & \textbf{lr} &  \textbf{time} & \textbf{acc.} & 1 & 2 & 3 & 5 & 10 & 20 & 30 & 50 & \textbf{MRR} \\\hline
    \multirow{12}{*}{3000} & \multirow{3}{*}{GCN} 
    & 0.001 & 42s & 92.77 & 0.51 & 0.65 & 0.72 & 0.79 & 0.84 & 0.88 & 0.91 & 0.94 & 0.63 \\
    & & 0.01 & 37s & 94.32 & 0.61 & 0.75 & 0.80 & 0.85 & 0.90 & 0.94 & 0.96 & 0.97 & 0.72 \\
    & & 0.05 & 32s & 92.72 & 0.46 & 0.64 & 0.70 & 0.78 & 0.84 & 0.89 & 0.91 & 0.94 & 0.60 \\
    \cmidrule{2-14}
    & \multirow{3}{*}{ClusterGCN}
    & 0.001 & 64s & 91.82 & 0.35 & 0.54 & 0.64 & 0.74 & 0.81 & 0.88 & 0.90 & 0.93 & 0.52 \\
    & & 0.01 (2 runs) & 86s & 91.34 & 0.44 & 0.56 & 0.69 & 0.74 & 0.81 & 0.87 & 0.90 & 0.93 & 0.57 \\
    & & 0.05 \\
    \cmidrule{2-14}
    & \multirow{3}{*}{GAT}
    & 0.001 & 77s & 90.48 & 0.34 & 0.47 & 0.55 & 0.67 & 0.77 & 0.84 & 0.87 & 0.90 & 0.48 \\
    & & 0.01 (2 runs) & 76s & 91.01 & 0.38 & 0.50 & 0.56 & 0.65 & 0.77 & 0.84 & 0.87 & 0.91 & 0.50 \\
    & & 0.05 \\
    \cmidrule{2-14}
    & \multirow{3}{*}{SAGE}
    & 0.001 & 59s & 93.19 & 0.42 & 0.58 & 0.62 & 0.73 & 0.83 & 0.89 & 0.92 & 0.94 & 0.56 \\
    & & 0.01 & 61s & 96.09 & 0.61 & 0.75 & 0.80 & 0.87 & 0.92 & 0.96 & 0.97 & 0.98 & 0.72 \\
    & & 0.05 & 46s & 96.22 & 0.65 & 0.77 & 0.83 & 0.87 & 0.93 & 0.96 & 0.97 & 0.98 & 0.75 \\    
    \midrule
    \multirow{12}{*}{10000} & \multirow{3}{*}{GCN}
    & 0.001 & 386s & 95.95 & 0.58 & 0.67 & 0.72 & 0.79 & 0.86 & 0.91 & 0.93 & 0.97 & 0.67 \\
    & & 0.01 & 213s & 90.91 & 0.45 & 0.60 & 0.63 & 0.67 & 0.74 & 0.80 & 0.83 & 0.86 & 0.56 \\
    & & 0.05 (1 run) & 172s & 91.89 & 0.47 & 0.52 & 0.54 & 0.62 & 0.68 & 0.74 & 0.78 & 0.83 & 0.54 \\
    \cmidrule{2-14}
    & \multirow{3}{*}{ClusterGCN}
    & 0.001 & 359s & 90.90 & 0.31 & 0.45 & 0.55 & 0.60 & 0.68 & 0.76 & 0.80 & 0.83 & 0.44 \\
    & & 0.01 (2 runs) & 338s & 86.21 & 0.49 & 0.54 & 0.58 & 0.62 & 0.68 & 0.73 & 0.75 & 0.77 & 0.55 \\
    & & 0.05 \\
    \cmidrule{2-14}
    & \multirow{3}{*}{GAT}
    & 0.001 & 351s & 89.68 & 0.35 & 0.41 & 0.45 & 0.54 & 0.63 & 0.72 & 0.75 & 0.79 & 0.44 \\
    & & 0.01 & 322s & 87.70 & 0.31 & 0.45 & 0.53 & 0.58 & 0.67 & 0.72 & 0.74 & 0.80 & 0.44 \\
    & & 0.05 (2 runs) & 109s & 71.24 & 0.33 & 0.37 & 0.39 & 0.41 & 0.44 & 0.47 & 0.49 & 0.51 & 0.37 \\
    \cmidrule{2-14}
    & \multirow{3}{*}{SAGE}
    & 0.001 & 357s & 96.08 & 0.54 & 0.67 & 0.73 & 0.80 & 0.86 & 0.91 & 0.93 & 0.95 & 0.65 \\
    & & 0.01 & 283s & 96.65 & 0.60 & 0.71 & 0.76 & 0.82 & 0.88 & 0.92 & 0.94 & 0.96 & 0.70 \\
    & & 0.05 & 267s & 97.59 & 0.60 & 0.72 & 0.82 & 0.87 & 0.92 & 0.95 & 0.96 & 0.98 & 0.72 \\
    \bottomrule
\end{tabular}
\end{adjustbox}
\caption{\label{tab:gnntype}Tests on \sampleu with 50 training epochs on 5-layer architectures, with \purity{80}{10} as both the train and test purity (average of three runs).  The empty rows correspond to meaningless results (no learning is actually taking place).}
\end{table}

On the graph with 3000 nodes, GCN performance is comparable with SAGE, even slightly better regarding some measures; on the other hand, ClusterGCN and GAT perform slightly worse and are not able to learn for large values of the learning rate.
On the graph with 10000 nodes, SAGE performs best and, in addition, is more flexible since it provides meaningful results for all values of the learning rate.

In the following, SAGE will be used as the reference GNN architecture.

\subsubsection{Comparison on the type of graph}

Table \ref{tab:graphtype3000} compares the model on equal-sized graphs (all with 3000 nodes, or around 3000 in the case of \ego)  with different structure. 
Results are the average of five runs: in the case of \ego, \ba, \er, and \sampleba, which are randomly generated, this implies that each run uses a different graph. The (non-constant) average number of nodes in \ego is specified as a subscript in the tables.


\noindent
\begin{table}[h!]
\begin{adjustbox}{max width=\textwidth}
\begin{tabular}{cccccccccccccc}
    \toprule
    & \textbf{train} & \% & \multicolumn{8}{c}{\texttt{hits@k} (out of 600 neg. examples)} & \\
    \textbf{graph} & \textbf{time} & \textbf{acc.} & 1 & 2 & 3 & 5 & 10 & 20 & 30 & 50 & \textbf{MRR} \\\hline
    \sample & 52s & 80.22 & 0.04 & 0.08 & 0.11 & 0.17 & 0.27 & 0.40 & 0.50 & 0.64 & 0.12 \\
    \sampleu & 57s & 91.69 & 0.35 & 0.46 & 0.56 & 0.67 & 0.77 & 0.85 & 0.88 & 0.91 & 0.49 \\
    $\ego_{3389}$ & 61s & 81.70 & 0.26 & 0.33 & 0.38 & 0.44 & 0.54 & 0.67 & 0.76 & 0.86 & 0.35 \\
    \ba & 49s & 70.25 & 0.04 & 0.07 & 0.09 & 0.12 & 0.19 & 0.28 & 0.34 & 0.44 & 0.09 \\
    \er & 40s & 70.99 & 0.05 & 0.07 & 0.10 & 0.13 & 0.20 & 0.30 & 0.38 & 0.47 & 0.10 \\
    \sampleba & 37s & 55.28 & 0.03 & 0.05 & 0.07 & 0.11 & 0.15 & 0.21 & 0.24 & 0.28 & 0.07 \\
     \bottomrule
\end{tabular}
\end{adjustbox}
\caption{\label{tab:graphtype3000}Tests on a 5-layer SAGE architecture, with \purity{80}{10} as both the train and test purity.}
\end{table}

Table \ref{tab:graphtype10000} does the same on graphs with 10000 nodes.

\noindent
\begin{table}[h!]
\begin{adjustbox}{max width=\textwidth}
\begin{tabular}{cccccccccccccc}
    \toprule
    & \textbf{train} & \% & \multicolumn{8}{c}{\texttt{hits@k} (out of 2000 neg. examples)} & \\
    \textbf{graph} & \textbf{time} & \textbf{acc.} & 1 & 2 & 3 & 5 & 10 & 20 & 30 & 50 & \textbf{MRR} \\\hline
    \sample & 357s & 81.99 & 0.04 & 0.07 & 0.10 & 0.14 & 0.21 & 0.30 & 0.37 & 0.47 & 0.10 \\
    \sampleu & 321s & 94.98 & 0.41 & 0.57 & 0.63 & 0.72 & 0.81 & 0.87 & 0.9 & 0.92 & 0.55 \\
    $\ego_{10010}$ & 278s & 89.91 & 0.25 & 0.36 & 0.41 & 0.48 & 0.6 & 0.69 & 0.75 & 0.81 & 0.37 \\
    \ba & 301 & 72.60 & 0.04 & 0.06 & 0.08 & 0.11 & 0.16 & 0.23 & 0.29 & 0.38 & 0.08 \\
    \er & 271 & 72.08 & 0.03 & 0.05 & 0.07 & 0.09 & 0.14 & 0.22 & 0.27 & 0.35 & 0.07 \\
    \sampleba & 290 & 72.19 & 0.06 & 0.09 & 0.11 & 0.13 & 0.17 & 0.21 & 0.24 & 0.28 & 0.10 \\
    \bottomrule
\end{tabular}
\end{adjustbox}
\caption{\label{tab:graphtype10000}Tests on a 5-layer SAGE architecture, with \purity{80}{10} as both the train and test purity.}
\end{table}

\subsubsection{Comparison on graph size}

Table \ref{tab:size} shows the training time and the test results on different graph sizes for both \sample and \sampleu, with \purity{80}{10} as both the purity of the train and the purity of the test.
The batch size at training time is adapted to the size of the graph.
Results for 3000 and 10000 nodes are the average of five runs, and results for 20000 nodes are the average of two runs.

\noindent
\begin{table}[h!]
\begin{adjustbox}{max width=\textwidth}
\begin{tabular}{cccccccccccccc}
    \toprule
    & & \textbf{train} & \% & \multicolumn{8}{c}{\texttt{hits@k} (out of 20\% neg. examples)} & \\
    \textbf{graph} & \textbf{\#nodes} & \textbf{time} & \textbf{acc.} & 1 & 2 & 3 & 5 & 10 & 20 & 30 & 50 & \textbf{MRR} \\\hline
    \multirow{4}{*}{\rotatebox[origin=c]{45}{$\sampleu$}}%
    & 3000 & 57s & 91.69 & 0.35 & 0.46 & 0.56 & 0.67 & 0.77 & 0.85 & 0.88 & 0.91 & 0.49 \\
    & 10000 & 321s & 94.98 & 0.41 & 0.57 & 0.63 & 0.72 & 0.81 & 0.87 & 0.9 & 0.92 & 0.55 \\
    & 20000 & 1001s & 96.67 & 0.42 & 0.63 & 0.72 & 0.79 & 0.85 & 0.9 & 0.92 & 0.94 & 0.59 \\
    & 50000 & 4261s & 88.82 & 0.34 & 0.42 & 0.45 & 0.50 & 0.53 & 0.58 & 0.62 & 0.68 & 0.41 \\
    \bottomrule
\end{tabular}
\end{adjustbox}
\caption{\label{tab:size}Tests with 50 epochs on a 5-layer SAGE architecture, with \purity{80}{10} as both the train and test purity.}
\end{table}

Hits@k results refer to a total number of negative examples equal to the number of test nodes (20\% of the total number of nodes).
Results indicate that there is no significant decrease of the precision as the graph grows bigger, even tough, due to the way \sampleu is computed (the $n$ nodes with the highest degree), the average node degree is smaller in bigger graphs (i.e., there tend to be less connections, which certainly does not help learning).
In fact, Hits@k and MRR results indicate that the algorithm scales fairly well because the total number of negative examples grows larger with the size of the graph, making it more and more difficult for a positive example to be ranked in the top $k$; however, MRR is still high with 50 thousand nodes.

Training times grow slightly more than linearly on the number of nodes; this is because (1) there is a computational overhead for having to deal with more data; and (2) early stopping occurs more often in smaller graphs.

\subsubsection{Comparison on the type of node features}

Table \ref{tab:features} compares two kinds of training (both on a graph with ten thousand nodes) that depend on the information passed to the GNN in form of \emph{node features}:
\begin{itemize}
    \item \texttt{original} means that the 128-dimension node features that are available in \ogblc are used;
    \item \texttt{dummy} means that those features have been replaced by a small vector of three constants $1$ (therefore, containing no information about nodes). 
\end{itemize}

\noindent
\begin{table}[h!]
\begin{adjustbox}{max width=\textwidth}
\begin{tabular}{cccccccccccccc}
    \toprule
    & \textbf{node} & \textbf{train} & \% & \multicolumn{8}{c}{\texttt{hits@k} (out of 3000 neg. examples)} & \\
    \textbf{graph} & \textbf{features} & \textbf{time} & \textbf{acc.} & 1 & 2 & 3 & 5 & 10 & 20 & 30 & 50 & \textbf{MRR} \\\hline
    \multirow{2}{*}{\rotatebox[origin=c]{10}{$\sample_{10K}$}}%
    & \texttt{original} 
    & 357s & 81.99 & 0.04 & 0.07 & 0.10 & 0.14 & 0.21 & 0.30 & 0.37 & 0.47 & 0.10 \\
    & \texttt{dummy}
    & 350s & 83.20 & 0.06 & 0.08 & 0.11 & 0.16 & 0.22 & 0.33 & 0.39 & 0.46 & 0.11 \\
    \midrule
    \multirow{2}{*}{\rotatebox[origin=c]{10}{$\sampleu_{10K}$}}%
    & \texttt{original} 
    & 321s & 94.98 & 0.41 & 0.57 & 0.63 & 0.72 & 0.81 & 0.87 & 0.90 & 0.92 & 0.55 \\
    & \texttt{dummy}
    & 355s & 94.68 & 0.41 & 0.58 & 0.65 & 0.72 & 0.79 & 0.84 & 0.87 & 0.90 & 0.55 \\
    \bottomrule
\end{tabular}
\end{adjustbox}
\caption{\label{tab:features}Tests with 50 epochs on a 5-layer SAGE architecture, with \purity{80}{10} as both the train and test purity.}
\end{table}

The results indicate that the information contained in the node features on the original node features does not play a role in this setting; this suggests that practically all the information that the model needs is already contained in the graph connectivity. However, this may not be the case with other graphs.

\subsubsection{Comparison on different values for train and test purity}
\label{sec:ComparisonOnDifferentValuesForTrainAndTestPurity}

Table \ref{tab:purity} compares different settings in which the purity of the training and test examples (Section \ref{fig:posNegExamples}) has been set to different values.

Each three lines of the table correspond to the average of five training runs with the specified purity for training examples.
Subsequently, for each run, test examples with different purity are generated after training has been performed, so that the same trained model corresponding to the first line is tested several times (first, second, and third line).
Moreover, we recall that each testing phase is, in turn, the average of five tasks where different sets of test examples have been randomly generated).

\noindent
\begin{table}[h!]
\begin{adjustbox}{max width=\textwidth}
\begin{tabular}{ccccccccccccccc}
    \toprule
    & \textbf{train} & \textbf{train} & \textbf{test} & \% & \multicolumn{8}{c}{\texttt{hits@k} (out of 3000 neg. examples)} & \\
    & \textbf{graph} & \textbf{purity} & \textbf{purity} & \textbf{time} & \textbf{acc.} & 1 & 2 & 3 & 5 & 10 & 20 & 30 & 50 & \textbf{MRR} \\\hline
    1 & \multirow{9}{*}{\rotatebox[origin=c]{75}{$\sample_{10K}$}}%
    & \multirow{3}{*}{\purity{100}{0}} & \purity{100}{0} & \multirow{3}{*}{350s}
    & 88.84 & 0.04 & 0.08 & 0.11 & 0.16 & 0.26 & 0.40 & 0.51 & 0.64 & 0.11 \\
    2 & & & \purity{80}{10} & 
    & 78.25 & 0.02 & 0.04 & 0.06 & 0.09 & 0.16 & 0.23 & 0.30 & 0.39 & 0.06 \\
    3 & & & \purity{50}{20} & 
    & 71.21 & 0.01 & 0.02 & 0.03 & 0.04 & 0.08 & 0.13 & 0.18 & 0.25 & 0.04 \\
    \cmidrule(lr){3-15} 
    4 & & \multirow{3}{*}{\purity{80}{10}} & \purity{100}{0} & \multirow{3}{*}{349s} 
    & 87.61 & 0.06 & 0.14 & 0.17 & 0.23 & 0.32 & 0.46 & 0.56 & 0.65 & 0.15 \\
    5 & & & \purity{80}{10} & 
    & 81.46 & 0.04 & 0.06 & 0.09 & 0.12 & 0.19 & 0.28 & 0.36 & 0.44 & 0.09 \\
    6 & & & \purity{50}{20} & 
    & 75.73 & 0.01 & 0.02 & 0.04 & 0.06 & 0.10 & 0.16 & 0.21 & 0.29 & 0.04 \\
    \cmidrule(lr){3-15} 
    7 & & \multirow{3}{*}{\purity{50}{20}} & \purity{100}{0} & \multirow{3}{*}{349s} 
    & 88.30 & 0.06 & 0.13 & 0.19 & 0.25 & 0.36 & 0.49 & 0.58 & 0.68 & 0.16 \\
    8 & & & \purity{80}{10} & 
    & 82.25 & 0.05 & 0.07 & 0.10 & 0.14 & 0.22 & 0.31 & 0.38 & 0.47 & 0.10 \\
    9 & & & \purity{50}{20} & 
    & 76.26 & 0.02 & 0.03 & 0.04 & 0.06 & 0.11 & 0.18 & 0.23 & 0.30 & 0.05 \\
    \midrule
    10 & \multirow{9}{*}{\rotatebox[origin=c]{75}{$\sampleu_{10K}$}}%
    & \multirow{3}{*}{\purity{100}{0}} & \purity{100}{0} & \multirow{3}{*}{356s} 
    & 98.50 & 0.74 & 0.89 & 0.92 & 0.96 & 0.98 & 0.98 & 0.99 & 0.99 & 0.84 \\
    11 & & & \purity{80}{10} & 
    & 91.39 & 0.47 & 0.60 & 0.65 & 0.71 & 0.77 & 0.81 & 0.83 & 0.86 & 0.58 \\
    12 & & & \purity{50}{20} & 
    & 84.97 & 0.17 & 0.26 & 0.37 & 0.45 & 0.55 & 0.62 & 0.66 & 0.71 & 0.29 \\
    \cmidrule(lr){3-15} 13 &
    & \multirow{3}{*}{\purity{80}{10}} & \purity{100}{0} & \multirow{3}{*}{349s} 
    & 98.52 & 0.83 & 0.91 & 0.94 & 0.96 & 0.97 & 0.99 & 0.99 & 0.99 & 0.89 \\
    14 & & & \purity{80}{10} & 
    & 95.96 & 0.47 & 0.64 & 0.71 & 0.78 & 0.86 & 0.90 & 0.92 & 0.94 & 0.61 \\
    15 & & & \purity{50}{20} & 
    & 90.89 & 0.24 & 0.33 & 0.38 & 0.48 & 0.60 & 0.71 & 0.76 & 0.81 & 0.35 \\
    \cmidrule(lr){3-15}
    16 & & \multirow{3}{*}{\purity{50}{20}} & \purity{100}{0} & \multirow{3}{*}{348s}
    & 98.23 & 0.84 & 0.91 & 0.92 & 0.94 & 0.96 & 0.97 & 0.98 & 0.99 & 0.89 \\
    17 & & & \purity{80}{10} &
    & 96.24 & 0.63 & 0.76 & 0.80 & 0.84 & 0.88 & 0.92 & 0.93 & 0.95 & 0.73 \\
    18 & & & \purity{50}{20} &
    & 91.73 & 0.26 & 0.38 & 0.46 & 0.56 & 0.66 & 0.75 & 0.79 & 0.83 & 0.39 \\
    \bottomrule
\end{tabular}
\end{adjustbox}
\caption{\label{tab:purity}Tests with 50 epochs on a 5-layer SAGE architecture (average of five runs).}
\end{table}

Not surprisingly, the results, both in terms of accuracy and MRR, get worse as soon as the test examples are \emph{noisier}, i.e., when test purity is different from $\purity{100}{0}$ (for example, consider lines 3, 6, 9, 12, 15, and 18 with respect to lines 1, 4, 7, 10, 13, and 16).

On the other hand, it is interesting to note that the test results are better when there is some \emph{noise} in the training data.
This suggests that the model is capable of \emph{ generalizing} from noisy data at training time, thus becoming better at classifying noisy data at test time.

\begin{example}
    Consider executions on \sampleu with test purity \purity{100}{0} (lines 10, 13, and 16). In this case, the accuracy results are basically similar for all values of the train purity (98.50\%, 98.52\%, and 98.24\%, respectively); however, MRR is better when there is noise in the training data (0.89 in lines 13 and 16, compared to 0.84 in line 10).
    
    The situation is similar when the purity of the test is \purity{80}{10} or even \purity{50}{20}: in this case, both accuracy and MRR improve clearly when the purity of the train is worse, that is, \purity{80}{10} or \purity{50}{20} (values 95.96\% and 96.24\% for accuracy in the case of \purity{80}{10}, respectively, compared to 91.39\% in the case of the purity of the train \purity{100}{0}; something similar happens with MRR: values 0.61 and 0.73 are obtained, respectively, compared to 0.58).
\end{example}

We believe that these interesting results show the ability of the model to learn the connections of new nodes even if the data are noisy or have been corrupted in some way.

\subsubsection{Comparison on the shape of test examples}
\label{sec:ComparisonOnTheShapeOfTestExamples}

Finally, Table \ref{tab:testExamples} sheds light on an issue about test examples.
One could think that the trained model is only able to correctly classify examples if some (large) proportion of training nodes is involved (remember that in a test example $\myexample{S}{t}$, $t$ has to be a test node, but $S$ could contain any kind of nodes); this would mean that results depend too much on a part of the graph that has already been seen at training time.

\noindent
\begin{table}[h!]
\begin{adjustbox}{max width=\textwidth}
\begin{tabular}{c|cccccccccccc}
\toprule
& 0-4 & 5-9 & 10-14 & 15-19 & 20-24 & 25+ \\ \midrule 0-4 & 91.72\% & 90.00\% & & & & & \\ & \scriptsize{(919/1002)} & \scriptsize{(9/10)} & & & & \\ 5-9 & 95.23\% & 97.55\% & & & & \\ & \scriptsize{(1636/1718)} & \scriptsize{(239/245)} & & & & & \\ 10-14 & 95.77\% & 97.77\% & 100.00\% & & & \\ & \scriptsize{(2195/2292)} & \scriptsize{(527/539)} & \scriptsize{(4/4)} & & & \\ 15-19 & 97.82\% & 97.64\% & 100.00\% & & & \\ & \scriptsize{(985/1007)} & \scriptsize{(331/339)} & \scriptsize{(4/4)} & & & \\ 20-24 & 99.03\% & 100.00\% & 100.00\% & & & \\ & \scriptsize{(205/207)} & \scriptsize{(143/143)} & \scriptsize{(15/15)} & & & \\ 25-29 & 100.00\% & 97.59\% & 100.00\% & & & \\ & \scriptsize{(52/52)} & \scriptsize{(81/83)} & \scriptsize{(15/15)} & & & \\ 30-34 & 100.00\% & 100.00\% & 100.00\% & 100.00\% & & \\ & \scriptsize{(21/21)} & \scriptsize{(59/59)} & \scriptsize{(25/25)} & \scriptsize{(4/4)} & & \\ 35-39 & 100.00\% & 100.00\% & 100.00\% & 100.00\% & & \\ & \scriptsize{(3/3)} & \scriptsize{(24/24)} & \scriptsize{(18/18)} & \scriptsize{(2/2)} & & \\ 40-44 & 100.00\% & 100.00\% & 100.00\% & 100.00\% & & \\ & \scriptsize{(4/4)} & \scriptsize{(18/18)} & \scriptsize{(16/16)} & \scriptsize{(8/8)} & & \\ 45-49 & & 100.00\% & 100.00\% & 100.00\% & 100.00\% & \\ & & \scriptsize{(11/11)} & \scriptsize{(16/16)} & \scriptsize{(11/11)} & \scriptsize{(1/1)} & \\ 50+ & & 100.00\% & 100.00\% & 100.00\% & 100.00\% & 100.00\% \\ & & \scriptsize{(7/7)} & \scriptsize{(26/26)} & \scriptsize{(29/29)} & \scriptsize{(12/12)} & \scriptsize{(10/10)} \\
\bottomrule
\end{tabular}
\end{adjustbox}
\caption{\label{tab:testExamples}Tests on \sampleu (10 thousand nodes) with 50 epochs on a 5-layer SAGE architecture. The results of two test executions (with different randomly generated test examples) after the same training are combined in this table. Globally, 7685 test examples out of 8000 have been correctly classified.}
\end{table}

However, Table \ref{tab:testExamples} shows that the distribution of nodes in test examples does not significantly affect the predictions.
Each row of the table corresponds to the number of training nodes involved in a test example (between 0 and 4, between 5 and 9, etc.), and each column corresponds to the number of test nodes involved in a test example (between 0 and 4, between 5 and 9, etc.). For instance, it is possible to see that, in this test, 2195 out of 2292 test examples have been correctly classified when the number of train nodes is between 15 and 19, and the number of test nodes is between 0 and 4 (third line, first column). The table shows that the accuracy values are similar on the diagonal and far away from it (very few examples are above the diagonal), so that it is possible to infer that test examples with a higher proportion of train nodes are classified similarly to examples with a higher proportion of test nodes. 


\subsection{Discussion}

Experiments demonstrate the feasibility of the proposed learning problem.
The first observation is that the model performs best on (the undirected version of) a realistic citation graph; it is not surprising that results are worse on artificially generated graphs like \ba or \er.

Another key observation is that the model benefits from learning on training data in which there is some amount of noise (represented by spurious edges, see Section \ref{sec:theProblemOfNewNodePrediction}). This aspect is unexpected and has to be verified by executing the proposed model on different realistic graphs.

For the Hits@k and MRR performance measures, the figures may be far from impressive in most tables.
However, it is important to note that both measures are heavily dependent on an implicit parameter, that is, the number of negative examples among which positive examples are supposed to be ranked.
In our setting, negative examples are usually in the thousands, so being in the top $k$ when $k$ is between $1$ and $50$ has to be considered satisfactory.

\section{Conclusions}
\label{con}

This paper contributes with the introduction and formalization of the task of ``new node prediction''. This problem presents two key differences with related tasks revised in the specialized literature: (1) the new node has no existing links from which to extract patterns for new predictions; and (2) the goal is to predict not just one, but all the links of this new node, or at least a significant part of them.

A deep learning architecture has been proposed for the new node prediction problem. The core of the architecture uses graph neural networks (GNNs). Experiments have used several GNN alternatives to instantiate the architecture, such as Graph convolutional network (GCN)  \cite{KipfWelling2017}, ClusterGCN \cite{ChiangEtAl2019}, Graph Attention Network (GAT)  \cite{VelickovicEtAl2018}, and GraphSAGE \cite{HamiltonEtAl2017}.

Experimental results with several synthetic graphs and with a subset of the \ogblc graph demonstrate the feasibility of solving this task. In addition, the experiments in this novel problem leave several observations of interest. Among others, experiments indicate that the proposed architecture benefits from using noisy training data. Experiments over this graph have also shown that learning is possible even without taking into account the features of the nodes, suggesting that the connections in the graph or topology provide most of the information in the dataset.

Finally, the trained architecture for the bibliographic citation network can be used as a bibliography recommendation system. An author can write a title and abstract of a paper, generate its embedding as node features, and solve the ``node prediction'' task to score sets of papers that can be considered as related work. 

Regarding future works,  we believe that this novel learning task can be used to gain insights into a variety of graphs; such as Knowledge Graphs (learning all relations of an unknown entity), social networks (learning friends of a new user), or protein-protein interaction networks (learning all known interactions of a protein). Therefore, our first future work is to explore the problem introduced and the architecture proposed in new data. The study of different graphs will lead to specific design choices. The second major future line is the hyperparameter optimization of the proposed architecture and the exploration of different neural architectures such as transformers.

 \bibliographystyle{abbrv}
 \bibliography{biblio}





\end{document}